\documentclass[12pt]{article}
\usepackage{bm,amssymb}
\usepackage{hyperref}
\usepackage[pdftex]{graphicx}

\newcommand{\ga}{\gamma}
\newcommand{\om}{\omega}
\newcommand{\ep}{\varepsilon}
\newcommand{\dl}{\delta}
\newcommand{\Tr}{{\rm Tr}\,}
\newcommand{\bk}{{\bf k}}
\newcommand{\la}{{\lambda}}

\newtheorem{theorem}{Theorem}

\title{Incoherent Quantum Control}
\author{{Alexander Pechen\thanks{E-mail: apechen@princeton.edu} } and
Herschel Rabitz\thanks{E-mail: hrabitz@princeton.edu}\\
\\
Department of Chemistry, Princeton University,\\ Princeton, New Jersey 08544,
USA}\date{}

\begin{document}
\maketitle

\begin{abstract}
Conventional approaches for controlling open quantum systems use coherent
control which affects the system's evolution through the Hamiltonian part of
the dynamics. Such control, although being extremely efficient for a large
variety of problems, has limited capabilities, e.g., if the initial and
desired target states have density matrices with different spectra or if a
control field needs to be designed to optimally transfer different initial
states to the same target state. Recent research works suggest extending
coherent control by including active manipulation of the non-unitary (i.e.,
incoherent) part of the evolution. This paper summarizes recent results specifically for
incoherent control by the environment (e.g., incoherent radiation or a gaseous
medium) with a kinematic description of controllability and landscape
analysis.
\end{abstract}

\section{Introduction}\label{sec1}
The manipulation of atomic or molecular quantum dynamics commonly uses
coherent quantum control, which may be extremely useful for a large variety of
problems~\cite{R,R0,PDR1987,R1,R2,RZ,R3,R4,R5,R6,R7,R8}. The dynamical
evolution of a closed quantum system under the action of a collection of
coherent controls $u=\{u_l(t)\}$ (e.g., Rabi frequencies of the applied laser
field) is described by the equation
\begin{equation}
\frac{d\rho_t}{dt}=-i\Bigl[H_0+\sum_l Q_l
u_l(t),\rho_t\Bigr],\qquad\rho|_{t=0}=\rho_0\label{eq1}
\end{equation}
Here $\rho_t$ is the system density matrix at time $t$ (for an $n$-level
quantum system, the set of all density matrices is ${\cal D}_n=\{\rho\in{\cal
M}_n\,|\, \rho\ge 0, \Tr\rho=1\}$, where ${\cal M}_n=\mathbb C^{n\times n}$ is
the set of $n\times n$ complex matrices), $H_0$ is the free system Hamiltonian
describing evolution of the system in the absence of control fields and each
$Q_l$ is an operator describing the coupling of the system to the control
field $u_l(t)$.

Coherent control of a closed system induces a unitary transformation of the
system density matrix $\rho_t=U_t\rho_0 U^\dagger_t$ and may have some
limitations. The first limitation is due to the fact that unitary
transformations of an operator preserve its spectrum; thus the spectrum of
$\rho_t$ is the same at any $t$ and, for example, a mixed state $\rho_0$ will
always remain mixed~\cite{schirmer2002}. A second limitation is that a control
$u_{\rm opt}$ which is optimal for some initial state $\rho_0$ may be not
optimal for another initial state $\tilde\rho_0$ even if $\rho_0$ and
$\tilde\rho_0$ have the same spectrum. This limitation originates from the
reversibility of unitary evolution and is due to the fact that $U_t\rho_0
U^\dagger_t\ne U_t\tilde\rho_0 U^\dagger_t$ if $\rho_0\ne\tilde\rho_0$. To
overcame these limitations at least to some degree, control by
measurements~\cite{qm1,qm2,roa1,pechen06_2,feng07} or incoherent
control~\cite{pechen06,Romano,acim,ding07} may be used, and in this work
incoherent control by the environment~\cite{pechen06} (ICE) is discussed. Some
general mathematical notions for the controlled quantum Markov dynamics are
formulated in Ref.~\cite{belavkin}.

The necessity to consider incoherent control relies also on the fact that
coherent control of quantum systems (e.g., of chemical reactions) in the
laboratory is often realized in a medium (solvent) which interacts with the
controlled system and plays the role of the environment. Furthermore, then
environment may be also affected to some degree by the coherent laser field,
thus effectively realizing incoherent control of the system. Moreover, laser
sources of coherent radiation at the present time have practical limitations,
and some frequencies are very expensive to generate compared to the respective
sources of incoherent control (e.g., incoherent radiation as considered in
Sec.~\ref{sec2.1} of this work). Thus the latter incoherent control can be
used in some cases to reduce the total cost of quantum control.

This paper summarizes recent results specifically for incoherent control by the environment\cite{pechen06} (ICE). A general theoretical formulation for incoherent control is provided in Sec.~\ref{sec2}, followed by the examples of control by incoherent radiation (Sec.~\ref{sec2.1}) and control through collisions with particles of a medium (e.g.,
solvent, gas, etc., Sec.~\ref{sec2.2}). Relevant known results about controllability and the structure of control landscapes for open quantum systems in the kinematic picture are briefly outlined in Sec.~\ref{sec:kraus}.

\section{Incoherent control by the environment}\label{sec2}
The dynamical evolution of an open quantum system under the action of coherent
controls in the Markovian regime is described by a master equation
\begin{equation}
\frac{d\rho_t}{dt}=-i\Bigl[H_0+H_{\rm eff}+\sum_l Q_l u_l(t),
\rho_t\Bigr]+{\cal L}\rho_t\label{eq2}
\end{equation}
The interaction with the environment modifies the Hamiltonian part of the
dynamics by adding an effective Hamiltonian term $H_{\rm eff}$ to the free
Hamiltonian $H_0$. Another important effect of the environment is the
appearance of the term $\cal L$ which describes non-unitary aspects of the
evolution and is responsible for decoherence. This term in the Markovian
regime has the general Gorini-Kossakowski-Sudarshan-Lindblad (GKSL)
form~\cite{GKS,lindblad}
\[
{\cal L}\rho=\sum\limits_i\left(2 L_i\rho L^\dagger_i-L^\dagger_i L_i\rho-\rho
L^\dagger_i L_i\right)
\]
where $L_i$ are some operators acting in the system Hilbert space. The
explicit form of the GKSL term depends on the particular type of the
environment, on the details of the microscopic interaction between the system
and the environment, and on the state of the environment.

The coherent portion of the control in~(\ref{eq2}) addresses only the
Hamiltonian part of the evolution while the GKSL part $\cal L$ remains fixed
(for the analysis of controllability properties for Markovian master equations
under coherent controls see for example, Ref.~\cite{altafini03}). However, the
generator $\cal L$ can also be controlled to some degree. For a fixed
system-environmental interaction, the generator $\cal L$ depends on the state
of the environment, which can be either a thermal state at some temperature
(including the zero temperature vacuum state) or an arbitrary non-equilibrium
state. Such a state is characterized by a (possibly, time dependent)
distribution of particles of the environment over their degrees of freedom,
which are typically the momentum $\bk\in\mathbb R^3$ and the internal energy
levels parameterized by some discrete index $\alpha\in A$ (e.g., for photons
$\alpha=1,2$ denotes polarization, for a gas of $N$-level particles
$\alpha=1,\dots,N$ denotes the internal energy levels). Denoting the density
at time $t$ of the environmental particles with momentum $\bk$ and occupying
an internal level $\alpha$ by $n_{\bk,\alpha}(t)$, and the corresponding GKSL
generator as ${\cal L}={\cal L}[n_{\bk,\alpha}(t)]$, the equation~(\ref{eq2})
becomes
\begin{equation}
\frac{d\rho_t}{dt}=-i\Bigl[H_0+H_{\rm eff}+\sum_l Q_l
u_l(t),\rho_t\Bigr]+{\cal L}[n_{\bk,\alpha}(t)]\rho_t\label{eq3}
\end{equation}
Here both $u_l(t)$ and $n_{\bk,\alpha}(t)$ are used as the controls, and for
$n_{\bk,\alpha}(t)$ the optimization is done over $\bk,\alpha$ in a time
dependent fashion to obtain a desired outcome.

The solution of~(\ref{eq3}) with the initial condition $\rho|_{t=0}=\rho_0$
for each choice of controls $\{u_l(t)\}$ and $n_{\bk,\alpha}(t)$ can be
represented by a family $P_t\{(u_l),n_{\bk,\alpha}\}$, $t\ge 0$ of completely
positive (CP), trace preserving maps (see Sec.~\ref{sec:kraus} for the
explicit definitions) as
\begin{equation}
\rho_t=P_t\{(u_l),n_{\bk,\alpha}\}\rho_0\label{ice:eq4}
\end{equation}
In general, for time dependent controls this family forms not a semigroup but
a self-consistent two-parameter family of CP, trace preserving maps
$P_{t,\tau}\{(u_l),n_{\bk,\alpha}\}$, $t\ge\tau\ge 0$, where each
$P_{t,\tau}\{(u_l),n_{\bk,\alpha}\}$ represents the evolution from $\tau$ to
$t$.

The target functional, also called the {\it performance index}, describes a
property of the controlled system which should be minimized during the control
and commonly consists of the two terms:
\[
J[(u_l),n_{\bk,\alpha}]=J_1[(u_l),n_{\bk,\alpha}]+J_2[(u_l),n_{\bk,\alpha}]
\]
The term $J_1[(u_l),n_{\bk,\alpha}]$, called the {\it objective functional},
represents the physical system's property which we want to minimize. The term
$J_2[(u_l),n_{\bk,\alpha}]$, called the {\it cost functional}, represents the
penalty for the control fields.

The first general class of objective functionals appears in the problem of
minimizing the expectation value of some observable associated to the system
at a target time $T>0$. The system is assumed at the initial time $t=0$ to be
in the state $\rho_0$. Any observable characterizing the system (e.g., its energy, population of some level, etc.) is represented by some self-adjoint operator $O$ acting in the system Hilbert space, and the corresponding objective functional has the form
\begin{equation}
J_1[(u_l),n_{\bk,\alpha}]=\Tr[\rho_T(u_l,n_{\bk,\alpha})
O]\equiv\Tr[(P_T\{(u_l),n_{\bk,\alpha}\}\rho_0)O]\label{o1}
\end{equation}
Here $\rho_T(u_l,n_{\bk,\alpha})\equiv P_T\{(u_l), n_{\bk,\alpha}\}\rho_0$ is
the final density matrix of the system evolving from the initial state
$\rho_0$ under the action of $u_l$ and $n_{\bk,\alpha}$. The physical meaning of this objective functional is that it represents the average measured value of the observable $O$ at the final time $T$ when the system evolves from the initial state $\rho_0$ under the action of the controls $(u_l),n_{\bk,\alpha}$.

The second general class of objective functionals appears in the problem of
optimal state-to-state transfer. Suppose that initially the system is in a
state $\rho_0$ and the control goal is to steer the system at some target time
$T$ into some desired target state $\rho_{\rm target}$. In this case one seeks
controls $(u_l)$  and $n_{\bk,\alpha}$ which minimize the distance between the
states $\rho_T(u_l, n_{\bk,\alpha})$ and $\rho_{\rm target}$. The
corresponding objective functional has the form
\begin{equation}
J_1[(u_l),n_{\bk,\alpha}]=\|\rho_T(u_l, n_{\bk,\alpha})-\rho_{\rm
target}\|\equiv\|P_T\{(u_l),n_{\bk,\alpha}\}\rho_0-\rho_{\rm
target}\|\label{o2}
\end{equation}
where $\|\cdot\|$ is a suitable matrix norm. Usually the Hilbert-Schmidt norm
$\|A\|=\sqrt{\Tr A^\dagger A}$ can be used.

The third important class of objective functionals appears in the problem of
producing a desired target CP, trace preserving map $P_{\rm target}$. In this
case the objective functional has the form
\begin{equation}
J_1[(u_l),n_{\bk,\alpha}]=\|P_T\{(u_l),n_{\bk,\alpha}\}-P_{\rm
target}\|\label{o3}
\end{equation}
where $\|\cdot\|$ is a suitable norm in the space of all CP, trace preserving
maps. In particular, in conventional models of quantum computation the target
transformation $P_{\rm target}$ is a unitary gate (e.g., a phase $U_{\phi}$ or
Hadamard $U_{\mathbb H}$ gate, and for these examples $P_{\rm
target}=U_{\phi}$ or $P_{\rm target}=U_{\mathbb H}$,
respectively)~\cite{mgrace,mgrace2}. More general non-unitary target
transformations can arise [e.g., in quantum computing with mixed
states~\cite{tarasov} or for generating controls robust to variations of the
initial system's state~\cite{rong07} (see also Sec.~\ref{sec4})].

The cost functional $J_2$ can be chosen to have the form
\[
J_2[(u_l),n_{\bk,i}]=\sum_l\int_0^T dt\alpha_l(t)|u_l(t)|^2+\max\limits_{0\le
t\le T}\sum\limits_i\int d\bk\beta_i(\bk)n_{\bk,i}(t)
\]
Here each function $\alpha_l(t)\ge 0$ [resp., $\beta_i(\bk)\ge 0$] is a weight
describing the cost for the control $u_l$ at time $t$ (resp., for the density
of particles of the environment with momentum $\bk$ and occupying the internal
level $i$). The first term minimizes the energy of the optimal coherent
control. The second term minimizes the total density of the environment.

The control functions belong to some sets of admissible controls $(u_l)\in\cal
E$ and $n_{\bk,\alpha}\in\cal D$. The following three important problems
arise.

{\it Optimal controls.} Find, for a given initial state $\rho_0$ and a target
time $T$, some (or all) controls $(u_l)\in\cal E$ and $n_{\bk,\alpha}\in\cal
D$ which minimize the performance index.

{\it Reachable sets.} Find, for a given final time $T>0$ and an initial state
$\rho_0$, the set of all states reachable from $\rho_0$ up to the time $T$,
i.e., the set
\[
{\cal R}_T(\rho_0)=\{P_t\{(u_l), n_{\bk,\alpha}\}\rho_0\,|\,t\le T,
(u_l)\in{\cal E}, n_{\bk,\alpha}\in{\cal D}\}
\]

{\it Landscape analysis.} Find, for a given $T>0$, an initial state $\rho_0$
and a self-adjoint operator $O$, all extrema (global and local, and saddles,
if any) of the objective functional $J_1[u_l(t),n_{\bk,\alpha}(t)]=\Tr[\{P_T[(u_l),n]\rho_0\} O]$ defined by~(\ref{o1}) [and similarly for the objective functionals defined by~(\ref{o2}) and~(\ref{o3})] or of the corresponding performance index.

\subsection{Incoherent control by radiation}\label{sec2.1}
Non-equilibrium radiation is characterized by its distribution in photon
momenta and polarization. For control with distribution of incoherent
radiation the magnitude of the photon momentum $|\bk|$ can be exploited along
with the polarization and the propagation direction in cases where
polarization dependence or spatial anisotropy is important (e.g., for
controlling a system consisting of oriented molecules bound to a surface).

A thermal equilibrium distribution for photons at temperature $T$ is
characterized by Planck's distribution
\[
n_\bk=\frac{1}{\exp\left(\frac{c\hbar|\bk|}{k_{\rm B}T}\right)-1}
\]
where $c$ is the speed of light, $\hbar$ and $k_{\rm B}$ are the Planck and
the Boltzmann constants which we set to one below. Non-equilibrium incoherent
radiation may have a distribution given as an arbitrary non-negative function
$n_{\bk,\alpha}(t)$. Some practical means to produce non-equilibrium
distributions in the laboratory may be based either on filtering thermal
radiation or on the use of independent monochromatic sources.

The master equation for an atom or a molecule interacting with a coherent
electromagnetic field $E_c(t)$ and with incoherent radiation with a
distribution $n_\bk(t)$ in the Markovian regime has the form:
\begin{equation}\label{r:eq1}
\frac{d\rho_t}{dt}=-i[H_0+H_{\rm eff}-\mu E_c(t),\rho_t]+{\cal L}_{\rm
Rad}[n_\bk(t)]\rho_t
\end{equation}
The coherent part of the dynamics is generated by the free system's
Hamiltonian $H_0=\sum_n\ep_n P_n$ with eigenvalues $\ep_n$, forming the spectrum ${\rm spec}(H_0)$, and the corresponding projectors $P_n$, the effective Hamiltonian $H_{\rm eff}$
resulting from the interaction between the system and the incoherent
radiation, dipole moment $\mu$, and electromagnetic field $E_c(t)$.

The GKSL generator ${\cal L}={\cal L}_{\rm Rad}$ induced by the incoherent
radiation with distribution function $n_{\bk}(t)$ has the form (e.g., see
Ref.~\cite{AcLuVol})
\begin{equation}\label{eq5}
{\cal L}_{\rm Rad}[n_\bk(t)]\rho=
\sum\limits_{\om\in\Omega}[\ga^+_\om(t)+\ga^-_{-\om}(t)]
(2\mu_\om\rho\mu^\dagger_\om-\mu^\dagger_\om\mu_\om\rho-\rho\mu^\dagger_\om\mu_\om)
\end{equation}
Here the sum is taken over the set of all system transition frequencies
$\Omega=\{\ep_n-\ep_m\,|\,\ep_n,\ep_m\in{\rm spec}{H_0}\}$,
$\mu_\om=\sum\limits_{\ep_n-\ep_m=\om}P_m\mu P_n$, and the coefficients
\[
\ga^\pm_\om(t)=\pi\int d\bk\delta(|\bk|-\om)|g_\bk|^2[n_\bk(t)+(1\pm1)/2]
\]
determine the transition rates between energy levels with transition frequency
$\om$. The transition rates depend on the photon density $n_\bk(t)$. The
form-factor $g_\bk$ determines the coupling of the system to the $\bk$-th mode
of the radiation. Equation~(\ref{r:eq1}) together with the explicit
structure~(\ref{eq5}) of the GKSL generator provides the theoretical
formulation for analysis of control by incoherent radiation.

The numerical simulations illustrating the capabilities of learning control by
incoherent radiation to prepare prespecified mixed states from a pure state is
available~\cite{pechen06} along with a theoretical analysis of the set of
stationary states for the generator ${\cal L}_{\rm Rad}$ for some
models~\cite{acim}. Incoherent control by radiation can extend the
capabilities of coherent control by exciting transitions between the system's
energy levels for which laser sources are either unavailable at the present
time or very expensive compared with the corresponding sources of incoherent
radiation. Ref.~\cite{ding07} provides a simple experimental realization of
the combined coherent (by a laser) and incoherent (by incoherent radiation
emitted by a gas-discharge lamp) control of certain excitations in Kr atoms.

\subsection{Incoherent control by a gaseous medium}\label{sec2.2}
This section considers incoherent control of quantum systems through
collisions with particles of a surrounding medium (e.g., a gas or solvent of
electrons, atoms or molecules, etc.). This case also includes coherent control
of chemical reactions in solvents if the coherent field addresses not only the
controlled system but the solvent as well. The particles of the medium in this
treatment serve as the control and the explicit characteristic of the medium
exploited to minimize the performance index is in general a time dependent
distribution of the medium particles over their momenta $\bk$ and internal
energy levels $\alpha\in A$. This distribution is formally described by a
non-negative function $n:\mathbb R^3\times A\times \mathbb R\to\mathbb R_+$,
whose value $n_{\bk,\alpha}(t)$ (where $\bk\in\mathbb R^3, \alpha\in A$, and
$t\in\mathbb R_+$) has the physical meaning of the density at time $t$ of
particles of the surrounding medium with momentum $\bk$ and in internal energy
level $\alpha$. In this scheme one prepares a suitable, in general
non-equilibrium, distribution of the particles in the medium such that the
medium drives the system evolution through collisions in a desired way.

It may be difficult to practically create a desired non-equilibrium
distribution of medium particles over their momenta. In contrast, a
non-equilibrium distribution in the internal energy levels can be relatively
easily created, e.g., by lasers capable of exciting the internal levels of the
medium particles or through an electric discharge. Then the medium particles
can affect the controlled system through collisions and this influence will
typically depend on their distribution. A well known example of such control
is the preparation of population inversion in a He--Ne gas-discharge laser. In
this system an electric discharge passes through the He--Ne gas and brings the
He atoms into a non-equilibrium state of their internal degrees of freedom.
Then He--Ne collisions transfer the energy of the non-equilibrium state of the
He atoms into the high energy levels of the Ne atoms. This process creates a
population inversion in the Ne atoms and subsequent lasing. A steady electric
discharge can be used to keep the gas of helium atoms in a non-equilibrium
state to produce a CW He--Ne laser. This process can serve as an example of
incoherent control through collisions by considering the gas of He atoms as
the control environment (medium) and the Ne atoms as the system which we want
to steer to a desired (excited) state.

Quantum systems controlled through collisions with gas or medium particles in
certain regimes can be described by master equations with GKSL generators
whose explicit structure is different from the generator ${\cal L}_{\rm Rad}$
describing control by incoherent radiation. If the medium is sufficiently
dilute, such that the probability of simultaneous interaction of the control
system with two or more particles of the medium is negligible, then the
reduced dynamics of the system will be Markovian~\cite{dumcke,apv} and will be
determined by two body scattering events between the system and one particle
of the medium. Below we provide a formulation for control of quantum systems
by a dilute medium, although the assumption of diluteness is not a restriction
for ICE, and dense mediums might be used for control as well.

The master equation for a system interacting with coherent fields $u_l(t)$ and
with a dilute medium of particles with mass $m$ has the form~(\ref{eq3}) with
the generator ${\cal L}[n_{\bk,\alpha}(t)]={\cal L}_{\rm
Medium}[n_{\bk,\alpha}(t)]$ specified by the distribution function of the
medium $n_{\bk,\alpha}(t)$ and by the $T$-operator (transition matrix) for the
scattering of the system and a medium particle. Below we assume that the
particles of the medium are characterized only by their momenta and do not
have internal degrees of freedom; otherwise, the state of one particle of the
medium should have the form $|\bk,\alpha\rangle$, where $\alpha$ specifies the
internal degrees of freedom. A transition matrix element is
$T_{n,n'}(\bk,\bk')=\langle n,\bk|T|n',\bk'\rangle$, where
$|n,\bk\rangle\equiv|n\rangle|\bk\rangle$ denotes the product state of the
system discrete eigenstate $|n\rangle$ (an eigenstate of the system's free
Hamiltonian $H_0$ with eigenvalue $\ep_n$) and a translational state of the
system and a medium particle with relative momentum $\bk$. If the system is
fixed in space (we consider this case below corresponding to the system
particle being much more massive than the particles of the surrounding medium)
then $|\bk\rangle$ is a translation state of a medium particle. The general
case of relative system medium particle motion can be considered as well using
suitable master equations. We will use the notation $
T_\om(\bk,\bk'):=\sum\limits_{m,n:\,\,
\ep_m-\ep_n=\omega}T_{m,n}(\bk,\bk')|m\rangle\langle n|$. The density of
particles of the medium at momentum $\bk$ is denoted as $n_\bk(t)$, and the
set of all transition frequencies $\om$ of the system among the energy levels
of $H_0$ is denoted as $\Omega$. In this notation the GKSL generator is
\begin{eqnarray}
&&{\mathcal L}_{\rm Medium}[n_\bk(t)]\rho=2\pi\sum\limits_{\om\in\Omega}\int
d\bk
n_\bk(t)\int d\bk'\dl\left(\frac{|\bk'|^2}{2m}-\frac{|\bk|^2}{2m}+\om\right)\nonumber\\
&&\qquad\qquad\quad\times\Bigl[T_\om(\bk',\bk)\rho T^\dagger_\om(\bk',\bk)
-\frac{1}{2}\Bigl\{T^\dagger_\om(\bk',\bk)T_\om(\bk',\bk),\rho\Bigr\}\Bigr]\label{D2}
\end{eqnarray}
where $\{\cdot,\cdot\}$ denotes the anti-commutator. If the medium is at
equilibrium with inverse temperature $\beta$, then the density has the
stationary Boltzmann form $n_\bk(t)\equiv
n_\bk=C(\beta,n)\exp(-\beta|\bk|^2/2m)$. Here the normalization constant
$C(\beta,n)$ is determined by the condition $\int d\bk n_\bk=n$, where $n$ is
the total density of the medium. The structure of Eq.~(\ref{D2}) has been
discussed previously for equilibrium media~\cite{dumcke,apv} and for
non-equilibrium stationary media~\cite{p}. Non-equilibrium media may be
characterized by generally time dependent distributions. Equation~(\ref{eq3})
with ${\cal L}[n_\bk(t)]\equiv {\cal L}_{\rm Medium}[n_\bk(t)]$ provides the
general formulation for theoretical analysis of control by a coherent field
$u_l(t)$ and by a non-equilibrium medium with density $n_\bk(t)$.

\begin{figure}[t]\center
\includegraphics[scale=1]{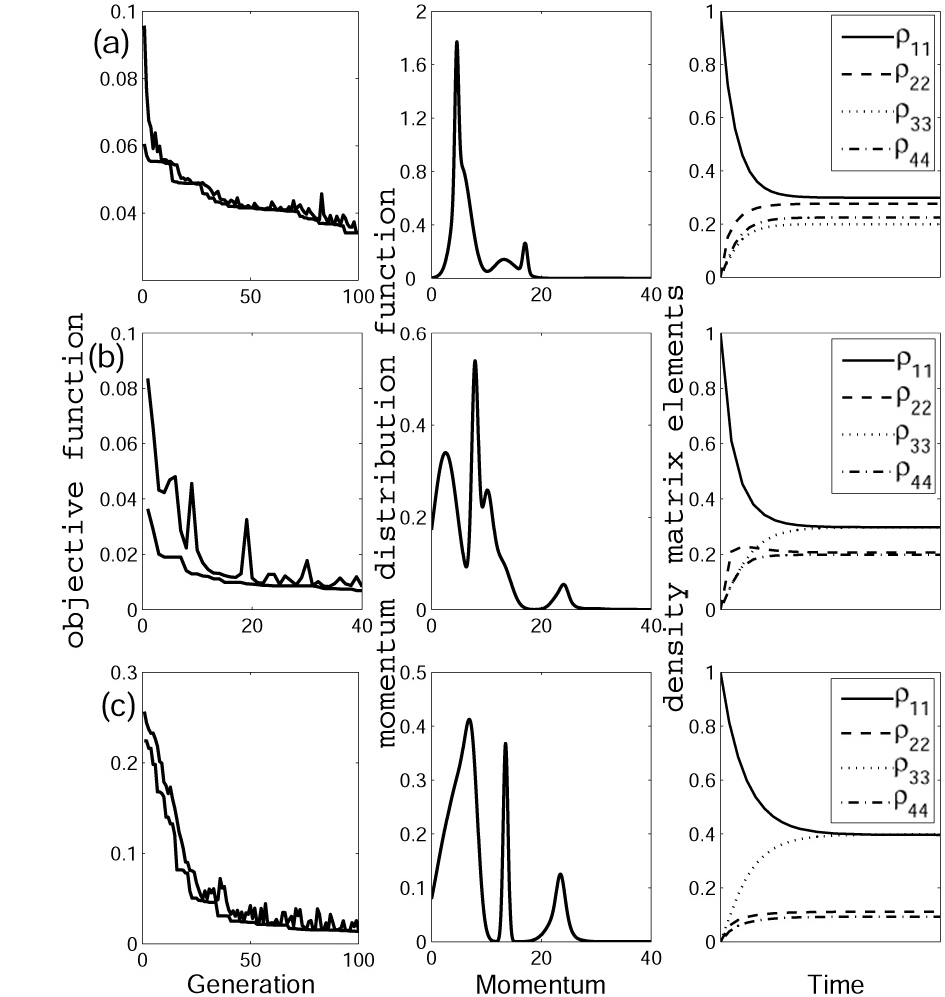}
\caption{(From Ref.~\cite{pechen06}. Copyright (2006) by the American Physical
Society.) Results of ICE simulations with a surrounding non-equilibrium medium
as the control for target states (a) $\rho_{\rm target}={\rm
diag}(0.3;\,0.3;\,0.2;\,0.2)$, (b) $\rho_{\rm target}={\rm
diag}(0.3;\,0.2;\,0.3;\,0.2)$, and (c) $\rho_{\rm target}={\rm
diag}(0.4;\,0.1;\,0.4;\,0.1)$. Each case shows: the objective function vs GA
generation, the optimal distribution vs momentum, and the evolution of the
diagonal elements of the density matrix for the optimal distribution. In the
plots for the objective function the upper curve is the average value for the
objective function and the lower one is the best value in each generation.}
\label{fig1ldl}
\end{figure}

As a simple illustration of such incoherent control, Fig.~\ref{fig1ldl}
reproduces the numerical results from Ref.~\cite{pechen06} for optimally
controlled transfer of a pure initial state of a four-level system into three
different mixed target states [i.e., the objective function~(\ref{o2}) is
chosen]. The control is modelled by collisions with a medium prepared in a
static non-equilibrium distribution $n_{|\bk|}$ whose form is optimized by
learning control using a genetic algorithm (GA)~\cite{GA} based on the
mutation and crossover operations. Since the initial and target states have
different spectra, they can not be connected by a unitary evolution induced by
coherent control. However, Fig.~\ref{fig1ldl} shows that ICE through
collisions can work perfectly for such situations.

\section{Kinematic description of incoherent control}\label{sec:kraus}
Physically admissible evolutions of an $n$-level quantum system can be
represented by CP, trace preserving maps ({\it Kraus maps})~\cite{kraus1983}.
A map $\Phi:{\cal M}_n\to{\cal M}_n$ is positive if for any $\rho\in{\cal
M}_n$ such that $\rho\ge 0$: $\Phi(\rho)\ge 0$. A linear map $\Phi:{\cal
M}_n\to{\cal M}_n$ is CP if for any $l\in\mathbb N$ the map
$\Phi\otimes\mathbb I_l:{\cal M}_n\otimes {\cal M}_l\to {\cal M}_n\otimes
{\cal M}_l$ is positive (here $\mathbb I_l$ denotes the identity map in ${\cal
M}_l$). A CP map $\Phi$ is called trace preserving if for any $\rho\in{\cal
M}_n:$ $\Tr\Phi(\rho)=\Tr\rho$. The conditions of trace preservation and
positivity for physically admissible evolutions are necessary to guarantee
that $\Phi$ maps states into states. The condition of complete positivity has
the following meaning. Consider the elements of ${\cal M}_l$ as operators of
some $l$-level ancilla system which does not evolve, i.e., its evolution is
represented by the identity mapping $\mathbb I_l$. Suppose that the $n$-level
system does not interact with the ancilla. Then the combined evolution of the
total system will be represented by the map $\Phi\otimes\mathbb I_l$ and the
condition of complete positivity requires that for any $l$ this map should
transform all states of the combined system into states, i.e. to be positive.

Any CP, trace preserving map $\Phi$ can be expressed using the Kraus
operator-sum representation as
\[
\Phi(\rho)=\sum\limits_{i=1}^\lambda K_i\rho K^\dagger_i
\]
where $K_i\in\mathbb C^{n\times n}$ are the Kraus operators subject to the
constraint $\sum_{i=1}^\lambda K^\dagger_i K_i=\mathbb I_n$ to guarantee trace
preservation. This constraint determines a complex Stiefel manifold
$V_n(\mathbb C^{\lambda n})$ whose points are $n\times (\lambda n)$ matrices
$V=(K_1;K_2;\dots;K_\lambda)$ (i.e., each $V$ is a column matrix of $K_1,\dots
K_\lambda$) satisfying the orthogonality condition $V^\dagger V=\mathbb I_n$.

The explicit evolution $P_t\{(u_l),n_{\bk,\alpha}\}$ in~(\ref{ice:eq4}) is
unlikely to be known for realistic systems. However, since this evolution is
always a CP, trace preserving map, it can be represented in the Kraus form
\begin{eqnarray*}
P_t\{(u_l),n_{\bk,\alpha}\}\rho&=&\sum_{i=1}^\la
K_i(t,(u_l),n_{\bk,\alpha})\rho K^\dagger_i(t,(u_l),n_{\bk,\alpha}),\\{\rm
where} &&\sum_{i=1}^\la
K^\dagger_i(t,(u_l),n_{\bk,\alpha})K_i(t,(u_l),n_{\bk,\alpha}\})=\mathbb I_n
\end{eqnarray*}
Assume that any Kraus map can be generated in this way using the available
coherent and incoherent controls $\{u_l(t)\}$ and $n_{\bk,\alpha}(t)$. Then
effectively the Kraus operators can be considered as the controls [instead of
$\{u_l(t)\}$ and $n_{\bk,\alpha}(t)$] which can be optimized to drive the
evolution of the system in a desired direction. This picture is called {\it
the kinematic picture} in contrast with {\it the dynamical picture} of
Sec.~\ref{sec2}. In the next two subsections we briefly outline the
controllability and landscape properties in the kinematic picture.

\subsection{Controllability}\label{sec4}
Any classical or quantum system at a given time is completely characterized by its state. The related notion of state controllability refers to the ability to steer the system from any initial state to any final state, either at a given time or asymptotically as time goes to infinity, and the important problem in control analysis is to establish the degree of state controllability for a given control system. Assuming for some finite-level system that the set of admissible dynamical controls generates arbitrary Kraus type evolution, the following  theorem implies then that the system is completely state controllable.
\begin{theorem} \label{t1}
For any state $\rho_{\rm f}\in{\cal D}_n$ of an $n$-level quantum system there
exists a Kraus map $\Phi_{\rho_{\rm f}}$ such that $\Phi_{\rho_{\rm
f}}(\rho)=\rho_{\mathrm{f}}$ for all states $\rho\in{\cal D}_n$.
\end{theorem}
{\bf Proof.} Consider the spectral decomposition of the final state $ \rho_{\mathrm{f}} =
\sum_{i=1}^n p_i |\phi_i \rangle\langle \phi_i|$, where $p_i$ is the
probability to find the system in the state $|\phi_i\rangle$ ($p_i \geq 0$ and
$\sum_{i=1}^n p_i = 1$). Choose an arbitrary orthonormal basis
$\{|\chi_j\rangle\}$ in the system Hilbert space and define the operators
\[
K_{ij}=\sqrt{p_i}\,|\phi_i\rangle\langle\chi_j|,\qquad i,j =1,\ldots,n.
\]
The operators $K_{i j}$ satisfy the normalization condition $\sum_{i,j=1}^n
K_{ij}^\dagger K_{ij}=\mathbb I_n$ and thus determine the Kraus map
$\Phi_{\rho_{\rm f}}(\rho) = \sum_{i,j=1}^n K_{i j} \rho K_{i j}^{\dagger}$.
The map $\Phi_{\rho_{\rm f}}$ acts on any state $\rho\in{\cal D}_n$ as
\[
\Phi_{\rho_{\rm f}}(\rho) = \sum_{i,j=1}^n p_i | \phi_i \rangle\langle \chi_j
| \rho |\chi_j \rangle\langle \phi_i | = ( \Tr \rho) \sum_{i=1}^n p_i | \phi_i
\rangle\langle \phi_i | = \rho_{\mathrm{f}}
\]
and thus satisfies the condition of the Theorem. $\Box$

The potential importance of this result is that it shows that there may exist
a single incoherent evolution which is capable for transferring all initial
states into a given target state, and moreover, the target state can be an
arbitrary pure or a mixed state\cite{rong07}. Thus this theorem shows that non-unitary
evolution can break the two general limitations for coherent unitary control
described in the second paragraph in the Introduction.

\subsection{Control landscape structure}
In the kinematic description, under the assumption that any Kraus map can be
generated, the objective functional becomes a function on the Stiefel manifold
$V_n(\mathbb C^{\lambda n})$. In practice, various gradient methods may be
used to minimize such an objective function. If the objective function has a
local minimum then gradient based optimization methods can be trapped in this
minimum and will not provide a true solution to the problem. For such an
objective function, if the algorithm stops in some minimum one can not be sure
that this minimum is global and therefore this solution may be not
satisfactory. This difficulty does not exist if {\it a priori} information
about absence of local minima for the objective function is available as
provided by the following theorem for a general class of objective functions
of the form $J_1[K_1,\dots,K_\lambda]=\Tr[(\sum_{i=1}^\la K_i\rho
K^\dagger_i)O]$ in the kinematic picture.
\begin{theorem} For any $n\in\mathbb N$,
$\rho\in{\cal D}_n$, and for any Hermitian $O\in{\cal M}_n$ the objective
function $J_1[K_1,\dots,K_\la]=\Tr[(\sum_{i=1}^\la K_i\rho K^\dagger_i)O]$ on
the Stiefel manifold $V_n(\mathbb C^{\la n})$ does not have local minima or
maxima; it has global minimum manifold, global maximum manifold, and possibly
saddles whose number and the explicit structure depend on the degeneracies of
$\rho$ and $O$.
\end{theorem}
The case $n=2$ has been considered in detail in Ref.~\cite{pechen07}, where
the global minimum, maximum, and saddle manifolds are explicitly described for
each type of initial state $\rho$. In particular, it is found that the
objective function $J_1$ for a non-degenerate target operator $O$ and for a
pure $\rho$ (i.e., such that $\rho^2=\rho$) does not have saddle manifolds;
for the completely mixed initial state $\rho=\frac{1}{2}\mathbb I$, $J_1$ has
one saddle manifold with the value of the objective function $J_{\rm
saddle}=1/2$; and for any partially mixed initial state $J_1$ has two saddle
manifolds corresponding to the values of the objective function $J_{\rm
saddle}^{\pm}=(1\pm\|{\bf w}\|)/2$, where ${\bf
w}=\Tr[\rho\boldsymbol{\sigma}]$ and
$\boldsymbol{\sigma}=(\sigma_x,\sigma_y,\sigma_z)$ is the vector of Pauli
matrices (the vector $\bf w$ is in the unit ball, $\|\bf w\|\le 1$ and this
vector characterizes the initial state as $\rho=\frac{1}{2}[\mathbb
I_2+\langle{\bf w},\boldsymbol{\sigma}\rangle]$). The case of arbitrary $n$ is
considered in Ref.~\cite{rebing07}.

\section{Conclusions} This paper outlines recent results for incoherent control of quantum systems through their interaction with an environment. A general formulation for incoherent control through GKSL dynamics is given, followed by examples of incoherent radiation and a gaseous medium serving as the incoherent control environments. The relevant known results on controllability of open quantum systems subject to arbitrary Kraus type dynamics, as well as properties of the corresponding control landscapes, are also discussed.

\section*{Acknowledgments} This work was supported by the NSF and ARO. A. Pechen
acknowledges also partial support from the RFFI 08-01-00727-a and thanks the organizers of
the ''28-th Conference on Quantum Probability and Related Topics''
(CIMAT-Guanajuato, Mexico, 2007) Prof. R.~Quezada Batalla and Prof. L.~Accardi
for the invitation to present a talk on the subject of this work.


\begin{thebibliography}{99}
\bibitem{R} A.~G.~Butkovskiy and Y.~I.~Samoilenko {\em Control of Quantum-Mechanical
Processes and Systems} (Nauka, Moscow, 1984);\\
A.~G.~Butkovskiy and Y.~I.~Samoilenko {\em Control of Quantum-Mechanical
Processes and Systems} (Kluwer, Dordrecht, 1990) (Engl. Transl.).
\bibitem{R0} D. Tannor and S.~A.~Rice, {\em J. Chem. Phys.}
\href{http://dx.doi.org/10.1063/1.449767}{{\bf 83}, 5013 (1985)}.
\bibitem{PDR1987} A. P. Pierce, M. A. Dahleh and H. Rabitz, {\em Phys. Rev. A}
\href{http://dx.doi.org/10.1103/PhysRevA.37.4950}{{\bf 37}, 4950 (1988)}.
\bibitem{R1} R.~S. Judson and H.~Rabitz, {\em Phys. Rev. Lett.}
\href{http://dx.doi.org/10.1103/PhysRevLett.68.1500}{{\bf 68}, 1500 (1992)}.
\bibitem{R2} W.~S. Warren, H.~Rabitz and M.~Dahleh, {\em Science} {\bf 259}, 1581 (1993).
\bibitem{RZ} S.~A.~Rice and M.~Zhao {\em Optical Control of Molecular Dynamics} (Wiley, New York, 2000).
\bibitem{R3} H.~Rabitz, R.~de Vivie-Riedle, M.~Motzkus and K.~Kompa, {\em Science} {\bf 288}, 824 (2000).
\bibitem{R4} M.~Shapiro and P.~Brumer {\em Principles of the Quantum Control of Molecular Processes}
(Wiley-Interscience, Hoboken, NJ, 2003).
\bibitem{R5} I. A. Walmsley and H. Rabitz, {\em Physics Today} {\bf 56}, 43 (2003).
\bibitem{R6} M.~Dantus and V.~V.~Lozovoy, {\em Chem. Rev.} {\bf 104}, 1813
(2004).
\bibitem{R7} L. Accardi, S. V. Kozyrev and A. N. Pechen, {\em QP--PQ: Quantum Probability and White Noise
Analysis} vol.~{\bf XIX} ed. L.~Accardi, M.~Ohya and N.~Watanabe (World Sci.
Pub. Co., Singapore), 1 (2006); {\it E-print}
http://xxx.lanl.gov/abs/quant-ph/0403100.
\bibitem{R8} D. D'Alessandro {\em Introduction to Quantum Control and
Dynamics} (Chapman and Hall, Boca Raton, 2007).
\bibitem{schirmer2002} S. G. Schirmer, A. I. Solomon and J. V. Leahy, {\em J. Phys. A: Math. Gen.}
\href{http://dx.doi.org/10.1088/0305-4470/35/18/309}{\textbf{35}, 4125
(2002)}.
\bibitem{qm1} R. Vilela Mendes and V. I. Man'ko, {\em Phys. Rev. A}
\href{http://dx.doi.org/10.1103/PhysRevA.67.053404}{{\bf 67}, 053404 (2003)}.
\bibitem{qm2} A. Mandilara and J. W. Clark, {\em Phys. Rev. A}
\href{http://dx.doi.org/10.1103/PhysRevA.71.013406}{{\bf 71}, 013406 (2005)}.
\bibitem{roa1} L. Roa, A. Delgado, M. L. Ladron de Guevara
and A. B. Klimov, {\em Phys. Rev. A}
\href{http://dx.doi.org/10.1103/PhysRevA.73.012322}{{\bf 73}, 012322 (2006)}.
\bibitem{pechen06_2} A. Pechen, N. Il'in, F. Shuang and H. Rabitz, {\em Phys. Rev. A}
\href{http://dx.doi.org/10.1103/PhysRevA.74.052102}{{\bf 74}, 052102
(2006)};\\
{\it E-print} http://xxx.lanl.gov/abs/quant-ph/0606187.
\bibitem{feng07} F. Shuang, A. Pechen, T.-S. Ho and H. Rabitz,
{\em J. Chem. Phys.} \href{http://dx.doi.org/10.1063/1.2711806}{{\bf 126}, 134303 (2007)};\\
{\it E-print} http://xxx.lanl.gov/abs/quant-ph/0609084.
\bibitem{pechen06} A. Pechen and H. Rabitz, {\em Phys. Rev. A}
\href{http://dx.doi.org/10.1103/PhysRevA.73.062102}{{\bf 73}, 062102
(2006)};\\
{\it E-print} http://xxx.lanl.gov/abs/quant-ph/0609097.
\bibitem{Romano} R. Romano and D. D'Alessandro, {\em Phys. Rev. A}
\href{http://dx.doi.org/10.1103/PhysRevA.73.022323}{{\bf 73}, 022323 (2006)}.
\bibitem{acim} L. Accardi and K. Imafuku, {\em QP--PQ: Quantum Probability and White Noise
Analysis} vol.~{\bf XIX} ed. L.~Accardi, M.~Ohya and N.~Watanabe (World Sci.
Pub. Co., Singapore), 28 (2006).
\bibitem{ding07} Y. Ding {\it et al}, {\em Rev. Sci. Instruments} {\bf 78}, 023103 (2007).
\bibitem{belavkin} V. P. Belavkin, {\em Automatia and Remote Control} {\bf 44}, 178 (1983);\\
{\it E-print} http://arxiv.org/abs/quant-ph/0408003.
\bibitem{GKS} V. Gorini, A. Kossakowski and E.~C.~G.~Sudarshan, {\em J. Math. Phys.} {\bf 17}, 821
(1976).
\bibitem{lindblad} G. Lindblad, {\em Comm. Math. Phys.} {\bf 48}, 119 (1976).
\bibitem{altafini03} C. Altafini, {\em J. Math. Phys.} {\bf 44}, 2357 (2003).
\bibitem{mgrace} M. Grace, C. Brif, H. Rabitz, I.~A.~Walmsley, R.~L. Kosut and
D.~A.~Lidar, {\em J. Phys. B: At. Mol. Opt. Phys.} {\bf 40}, S103 (2007); {\it
E-print} http://xxx.lanl.gov/abs/quant-ph/0702147.
\bibitem{mgrace2} M. Grace, C. Brif, H. Rabitz, D. A. Lidar, I.~A. Walmsley and R.~L.
Kosut, {\it J. Modern Optics} {\bf 54}, 2339 (2007).
\bibitem{tarasov} V.~E. Tarasov, {\em J. Phys. A: Math. Gen.}
\href{http://dx.doi.org/10.1088/0305-4470/35/25/305}{{\bf 35}, 5207 (2002)}.
\bibitem{rong07} R. Wu, A. Pechen, C. Brif and H. Rabitz, {\em J. Phys. A: Math. Theor.}
\href{http://dx.doi.org/10.1088/1751-8113/40/21/015}{{\bf 40}, 5681 (2007)};\\
{\it E-print} http://xxx.lanl.gov/abs/quant-ph/0611215.
\bibitem{AcLuVol} L. Accardi, Y.~G. Lu and I.~V. Volovich {\em Quantum Theory and Its Stochastic Limit}
(Springer, Berlin, 2002).
\bibitem{dumcke} R. D\"umcke, {\em Comm. Math. Phys.} {\bf 97}, 331 (1985).
\bibitem{apv} L. Accardi, A.~N. Pechen and I.~V. Volovich, {\em Infin. Dimens. Anal. Quant. Probab. and Relat.
Topics} {\bf 6}, 431 (2003); {\it E-print}
http://xxx.lanl.gov/abs/math-ph/0206032.
\bibitem{p} A. Pechen, {\em QP-PQ: Quantum Probability and White Noise
Analysis} vol. {\bf XVIII} ed M. Sch\"urmann and U. Franz (World Sci. Pub.
Co., Singapore), 428 (2005);\\
{\it E-print} http://xxx.lanl.gov/abs/quant-ph/0607134.
\bibitem{GA} D. E. Goldberg {\em Genetic Algorithms in Search, Optimization
and Machine Learning} (Addison-Wesley, Reading, MA, 1989).
\bibitem{kraus1983} K. Kraus {\em States, Effects, and Operations} (Springer, Berlin, New York, 1983).
\bibitem{pechen07} A. Pechen, D. Prokhorenko, R. Wu and H. Rabitz {\em J. Phys. A: Math. Theor.}
\href{http://dx.doi.org/10.1088/1751-8113/41/4/045205}{{\bf 41}, 045205
(2008)}; {\it E-print} http://xxx.lanl.gov/abs/0710.0604.
\bibitem{rebing07} R. Wu, A. Pechen, H. Rabitz, M. Hsieh and B. Tsou, J. Math. Phys. 2008 (at press);\\
{\em E-print} http://xxx.lanl.gov/abs/0708.2119.
\end{thebibliography}
\end{document}